# Time-Of-Flight methodologies with large-area diamond detectors for ion characterization in laser-driven experiments

M. Salvadori [1,2,3], G. Di Giorgio[2], M. Cipriani[2], M. Scisciò[2], C. Verona[4], P.L. Andreoli[2], G. Cristofari[2], R. De Angelis[2], M. Pillon[2], N.E. Andreev[5], P. Antici[3], N. G. Borisenko[6], D. Giulietti[7,8], M. Migliorati[1,9], O. Rosmej[10,11], S. Zähter[10,11], F. Consoli[2]

[1] *University of Rome "La Sapienza", Piazzale Aldo Moro 5, Rome, Italy*

[2] *ENEA, Fusion and Nuclear Safety Department, C.R. Frascati, Italy*

[3] *INRS-EMT, Varennes, Québec, Canada*

[4] *Industrial Engineering Department, University of Rome "Tor Vergata"*

[5] *Joint Institute for High Temperatures, RAS, Izhorskaya st.13, Bldg. 2, 125412, Moscow, Russia*

[6] *P. N. Lebedev Physical Institute, RAS, Leninsky Prospekt 53, 119991, Moscow, Russia*

[7] *Department of Physics, University of Pisa, Largo Bruno Pontecorvo 3, Pisa, Italy*

[8] *INFN of Pisa, Largo Bruno Pontecorvo 3, Pisa, Italy*

[9] *INFN of Rome, Piazzale Aldo Moro 2, Rome, Italy*

[10] *GSI Helmotzzentrum für Schwerionenforschung GmbH, Planckstr. 1, Darmstadt, Germany*

[11] *Goethe University Frankfurt, Max-von-Laue-Str. 1, Frankfurt am Main, Germany*

**Abstract** Time-Of-Flight (TOF) technique coupled with semiconductor detectors is a powerful instrument to provide real-time characterization of ions accelerated because of laser-matter interactions. Nevertheless, the presence of strong electromagnetic pulses (EMPs) generated during the interactions, can severely hinder its employment. For this reason, the diagnostic system must be designed to have high EMP shielding. Here we present a new advanced prototype of detector, developed at ENEA-Centro Ricerche Frascati (Italy), with a large area (15 mm × 15 mm) polycrystalline diamond sensor having 150 µm thickness. The tailored detector design and testing ensure high sensitivity and, thanks to the fast temporal response, high energy resolution of the reconstructed ion spectrum. The detector was offline calibrated and then successfully tested during an experimental campaign carried out at the PHELIX laser facility ($E_L \sim 100$ J, $\tau_L = 750$ fs, $I_L \sim (1-2.5) \times 10^{19}$ W cm$^{-2}$) at GSI (Germany). The high rejection to EMP fields was demonstrated and suitable calibrated spectra of the accelerated protons were obtained.

___________________________________________________________________________

Correspondence to: Martina.salvadori@uniroma1.it, fabrizio.consoli@enea.it

# I. INTRODUCTION

Semiconductor detectors characterized by a wide band-gap, i.e. Silicon Carbide and Diamonds, are commonly used as time-resolved sensors for Time-Of-Flight (TOF) measurements [1–8]. This technique is a valuable instrument for the real-time characterization of charged particles accelerated during laser-plasma interaction [9]. In particular, when coupled to semiconductor detectors, it allows to retrieve the particle energy distribution with a good resolution, and to reconstruct the associated spectra with high accuracy [8,10]. The latter can be retrieved exploiting the working principle of semiconductor detectors. When ionizing radiation interacts with the bulk of the material, it releases its energy and produces a certain amount of free electron-hole pairs, accordingly [11]. These charges are then collected at the electrodes generating the signal. Knowing the detector response and the ion type, it is possible to retrieve the number of particles impinging onto the detector for each energy range [8,12].

The very high purity and crystalline quality of single crystal diamonds guarantee a considerably high value of charge carrier mobility and charge collection efficiency, resulting in TOF measurements characterized by a high temporal resolution, and thus allowing to retrieve spectra with remarkable energy resolution [8,11]. Nevertheless, single crystal diamonds can be produced only with small surfaces. In experiments of laser-matter interaction, this limits the solid angle of detection coverage, and thus the overall sensitivity of the detection system. Indeed there are several applications, for instance those characterized by low fluxes of emitted particles, where instead high sensitivities are required [13]. Also for the other cases, since the ion fluxes decrease with respect to energy, information on the maximum energy of the emitted particles can be retrieved accurately only by detectors with high sensitivity. For this purpose, a polycrystalline diamond structure of high quality offers the possibility to produce sensors characterized by a large-area, achieving high sensitivity while maintaining a good temporal resolution. In this work a prototype of a new detector developed at ENEA-Centro Ricerche Frascati (Italy), based on a commercial 150 µm thick high quality (II-a electronic grade) polycrystalline diamond structure, provided about 5 years ago by Diamond Detectors LTD (UK), is presented. The sensor, shown in Figure 1a, has a wide surface (15 mm × 15 mm). This allows to increase the overall detection area by about one order of magnitude with respect to the classical 5 mm x 5 mm monocrystalline detectors [5,7,8,12,14] covering a larger solid angle and enhancing the overall sensitivity of the diagnostic system. As shown in Figure 1b, the diamond wafer is enclosed between two metallic electrodes (Au) placed on its large surface which, when biased, provide a constant electric field throughout the diamond thickness, allowing for efficient charge collection and eventually ion beam characterization.

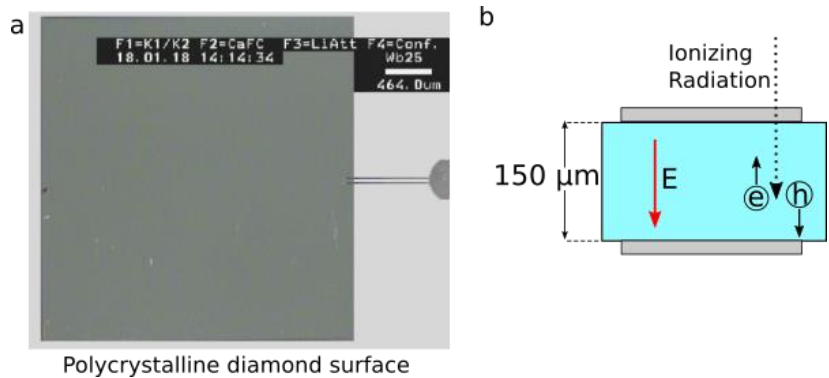

**Figure 1.** (a) A picture of the diamond surface taken with the Leica Wild M8 microscope equipped with the CCD camera JVC TK-C1480B. (b) A scheme of the diamond detector structure. The 150 µm polycrystalline diamond wafer is enclosed between two gold electrodes providing a constant electric field allowing for charge collection.

During efficient laser-plasma interactions, transient electromagnetic waves of remarkable intensity in the radiofrequency-microwave range are produced [15,16]. The presence of these strong electromagnetic pulses (EMPs) is a serious threat for the electronic devices placed near the interaction point. Indeed, for classical TOF ion diagnostics, EMPs can couple both with the detection system and the acquisition system often heavily hindering the effective measurement of accelerated ions [14]. To avoid this effect, an advanced time-of-flight procedure has already been developed [8,14]. This offers a good rejection to the EMPs and allows for the suitable spectral representation of the detected protons. The employment of a large-area sensor, on one side significantly improves the detector sensitivity, but on the other side makes the detection system more prone to coupling with EMPs. Particular attention has been thus dedicated to the design of an advanced housing of the sensor, with the purpose of achieving high rejection to EMPs. This will allow it to be fruitfully employed even in the typical harsh environments represented by high energy lasers interacting with matter, with the future purpose to employ this type of detectors successfully on the new facilities under development around the world, such as Apollon in France ($\tau_L = 15$ fs, $E_L = 150$ J, 10 PW) and L4 ATON in Czech Republic ($\tau_L = 150$ fs , $E_L = 1.5$ kJ, 10 PW).

In the following Sections we will first introduce the basic concept of the TOF technique coupled to semiconductor detectors. The novel detector design is then accurately described together with the performed characterization procedure. Eventually the results obtained during an experimental campaign where remarkable levels of EMPs were produced are presented and discussed.

## II. WORKING PRINCIPLE

**Time Of Flight technique and spectrum reconstruction.**

Time-Of-Flight technique relies on the measurement of the time required by a particle for traveling over a known distance $d_{TOF}$. Particles emitted from the source are collected by means of a time-resolved detector. The velocity distribution of the particles is thus transformed to a temporal distribution [17]. Since ionizing photons (UV-X) are also emitted during the interaction, the detection instant of the latter, $t_{ph}$, is used as an absolute reference, on the same oscilloscope trace to retrieve the actual interaction instant $t_{bang}$ [8], namely:

$$t_{bang} = t_{ph} - \frac{d_{TOF}}{c} \tag{1}$$

where $c$ is the speed of light in vacuum. The energy of the particles generating the signal can be then retrieved by using the relation:

$$E_i = m_i \, (\gamma_i - 1) \, c^2 \tag{2}$$

where $m_i$ is the ion mass and $\gamma_i$ the relativistic parameter.

Since the ion energies typically achieved in laser-plasma experiments are well below the relativistic limit, the previous relation simplifies to the classical $E_i \simeq \frac{1}{2} m_i v_i^2$.

Once the energy is known, from the amplitude of the signal, it is possible to obtain the number of particles impinging onto the detector by using the general relation [8,14]:

$$N_i = \frac{Q_c \epsilon_g}{q_e} \frac{1}{E_i CCE} \tag{3}$$

where $q_e$ is the electronic charge, and the radiation-ionization energy, $\epsilon_g$, is the average energy needed to create a free electron-hole pair inside the detector in use; for diamond detectors it is 13.1 eV [11]. CCE is the Charge Collection Efficiency of the detector [11] and $Q_c$ is the amount of collected charge that can be estimated by performing a numerical integration of the detected signal $V(t)$ [7,8]:

$$Q_c = k_A \int_{t_i}^{t_f} V(t) dt \tag{4}$$

Here, $k_A = \frac{A}{R}$, where $R$ is the impedance of the acquiring system and $A$ its attenuation. The time step to perform the integration is determined by the temporal resolution of the detection system, strongly dependent on the time response of the detector in use.

From the previous considerations it is therefore clear that for the spectrum estimation it is necessary to know the actual detector characteristics in terms of temporal response and Charge Collection Efficiency. To this purpose, the polycrystalline diamond was characterized by exposition to monochromatic α particles having $E_α = 5.486$ MeV, emitted by the $^{241}$Am radioactive source.

**Time response and Charge Collection Efficiency measurements.**

To appreciate the temporal response by single particle detection, the output of the diamond detector was connected to the biasing amplifier model DBA-IV (gain G = 46 dB, bandwidth = 2.5 GHz) and to the LeCroy 620 Zi scope (bandwidth = 2 GHz, sample rate = 20 GS/s). The fast amplifier was necessary to bias the detector and to increase the amplitude of the fast-varying signals to levels detectable by the scope.

The signal shown in Figure 2 has been obtained by averaging over several acquisitions performed with the detector biased at +300 V, typical bias used to feed the detector. The value of the applied voltage was chosen so to work in the velocity saturation regime, i.e. when the carriers velocity does not increase for a further increment of the suffered electric field [11].

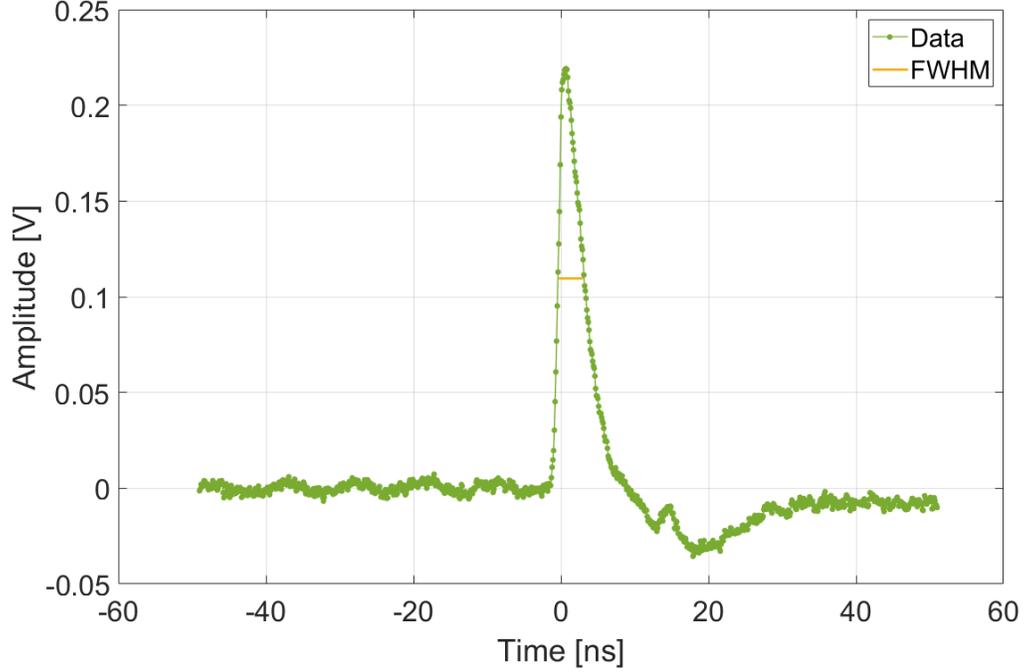

**Figure 2.** The signal provided by the polycrystalline diamond when exposed to single $\alpha$ particles of 5.486 MeV. This was obtained by averaging over ~ 100 acquisition and was used to retrieve both the temporal response of the sensor (3.47 ns) and its CCE (~45%).

From this kind of signal it is possible to retrieve the temporal response, by estimating the FWHM, and the charge collection efficiency. To estimate the latter it is first necessary to compute the number of charges ideally expected in the detector for a single incoming α particle, by exploiting the following relation [11]:

$$Q_g = \frac{q_e E_\alpha}{\epsilon_g} = 6.7 \times 10^{-14} \text{ C} \qquad (5)$$

Then, this value has to be compared to the one obtained during the measurement, which can be estimated by applying equation (4) this time using the multiplication factor $k_G = \frac{1}{RG}$ (being $G$ the gain of the amplifier) instead of $k_A$. The charge collection efficiency is thus evaluated by computing the ratio $\frac{Q_c}{Q_g}$. From the performed characterization the detector presents a FWHM and a CCE of 3.47 ns and ∼45% respectively.

**EMP shielding.**

To be effectively employed in environments characterized by high levels of EMPs, the diamond sensor was mounted in a case optimized for an optimal EMP rejection.

As shown in Figure 3a the diamond detector is located on a thin strip where all the electrical connection are made. In the developed detector design, these electrical connections are kept at a minimal length and are provided with an optimized grounding system and double-shielded cabling. The strip is then fixed on an ad-hoc support which allows its vertical alignment and to mount it onto a stainless steel cap (Figure 3b).

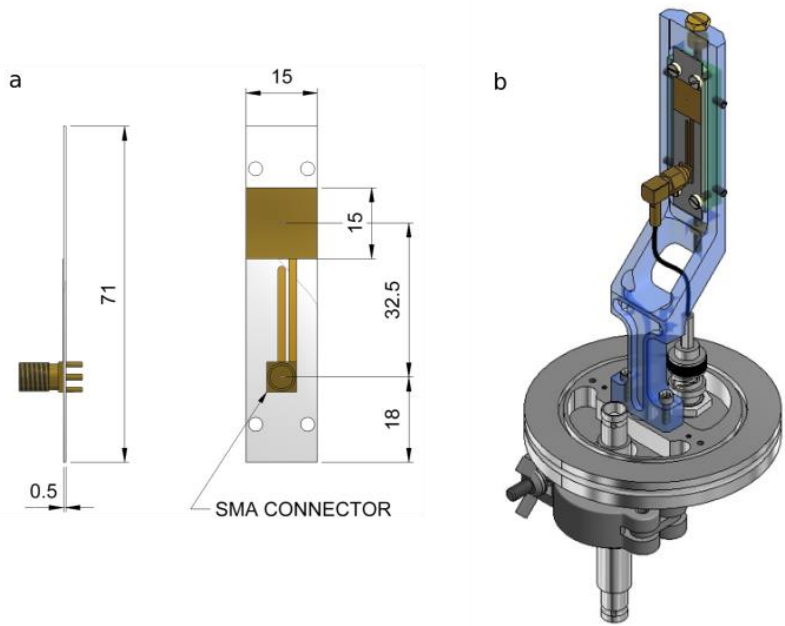

**Figure 3.** The inner detector layout: (a) The diamond wafer mounted on a thin strip together with the electrical connection, the bias to the electrodes is sent through an SMA connector. (b) The strip with the polycrystalline diamond detector is mounted on a support which allows its vertical alignment and is then fixed onto a stainless steel cap provided with proper feedthroughs.

The whole assembly is then inserted in the final holder. This consists in two cylindrical structures one inside the other. In Figure 4a and 4b, it is possible to see that the assembly hosting the diamond detector fits in the inner cylinder which, when properly closed, works as a Faraday cage providing an effective shielding to the detector from the external environment. The closure of the cage can be provided either with a metallic grid with proper characteristics or a thin metallic sheet.

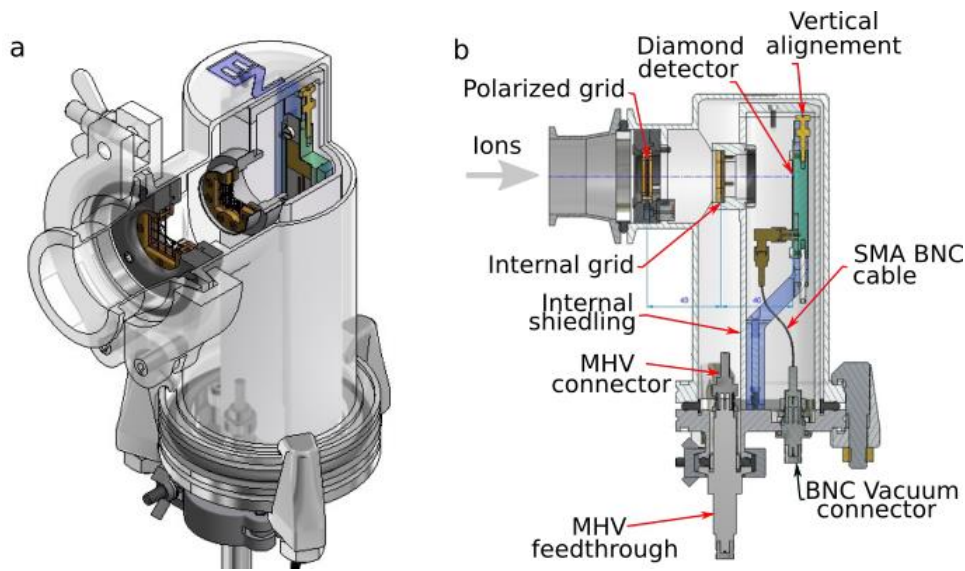

**Figure 4** (a) The external case in which the support shown in Figure 3b is mounted. It is possible to see the inner Faraday cage and the two grids (b) The scheme of the whole detector lay-out.

Indeed, the detector holder is designed to allow the positioning of two grids mounted parallel one to each other and to the diamond, as it is possible to see in Figure 4a and 4b. Both the grids are made of copper and the diameter of the wires used for the mesh is of 0.5 mm. The separation between the two grids is of 40 mm as well as the separation between the second grid and the diamond. The first grid, shown in Figure 5a, can be biased with voltages up to 5 kV through an internal cable with MHV connector. The step of the mesh is of 4.5 mm. This grid is isolated from the rest of the structure through a thermoplastic support made of white Delrin. The second internal grid, as shown in Figure 5b, is characterized by a denser wire mesh having a step of 2 mm. The grid material and geometrical parameters are chosen to be effectively used as closing component of the internal Faraday cage. Indeed, this grid is directly connected to the ground of the structure, and it is an integral part of the internal metallic shield, as it can be seen in Figure 4a and 4b. When the first grid is biased, the two grids assembly provides a static electric field which can be used to separate the electron contribution from the ionic component.

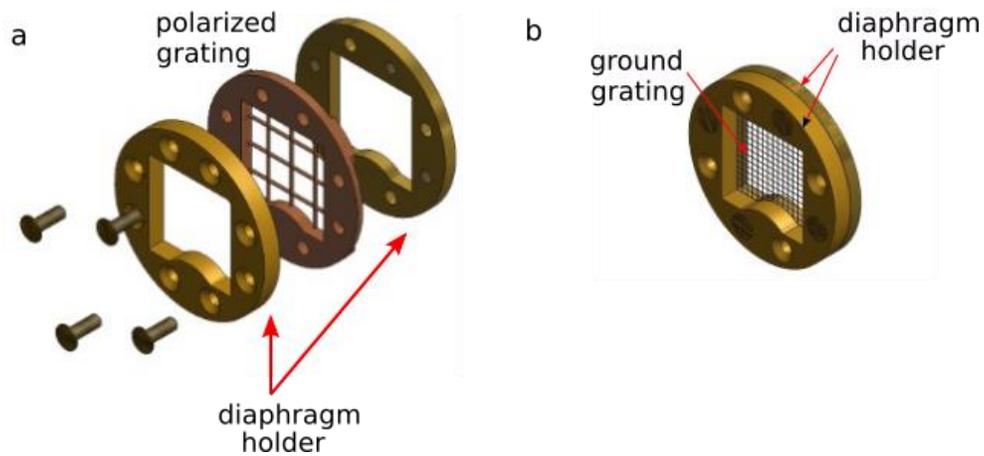

**Figure 5** the two copper grids placed in front of the diamond detector (a) The polarized grid can be biased up to 5 kV and has a mesh with a 4.5 mm step (b) The grounded grid is characterized by a denser mesh (2 mm step) which make it suitable also to close the inner Faraday cage.

The employment of these grids will reduce the actual active surface of the diamond by a factor ~20 %. Nevertheless, thanks to the large area of the polycrystalline diamond, this reduction does not remarkably affect the overall detector sensitivity and allows to use the detector also without any further conductive shield foil.

Nevertheless, the holder foresees the possibility to place a thin filter right in front of the diamond detector to cut heavier ion contribution, if necessary [8]. The combined action of the polarized grids and the filter, allows only protons to reach the detector leading to an easier spectrum reconstruction.

All these components are located in the external cylindrical case which is made in stainless steel with walls having 2 mm thickness. These walls provide an additional shielding for electromagnetic waves with frequency down to ∼ 500 Hz, exploiting the skin effect [18].

## III. EXPERIMENTAL RESULTS

The performances of the described detector were successfully tested during an experimental campaign carried out at the PHELIX laser facility (GSI, Germany) [19,20]. During the experiment two laser pulses were combined to study the effect of a controlled pre-plasma on laser-driven acceleration. A hydrodynamically stable, long-scale-length near-critical-density plasma was generated by irradiating low density polymer foams with a nanosecond pulse kept at an intensity of ∼ $5 \times 10^{13}$ W cm$^{-2}$. Then, with a delay of 2-3 ns, a short pulse of 750 fs was used to irradiate this plasma, delivering up to ∼100 J for an intensity of ∼ $(1 - 2.5) \times 10^{19}$ W cm$^{-2}$ [21]. This configuration including so short and energetic laser pulse is recognized to be one of the main scenarios where produced EMP fields are most intense [15]. Thus, it represented a very good test to prove the EMP rejection of the developed detector assembly.

The EMP level was monitored by means of two custom D-Dot differential electric-field probes [15]. During a typical shot, the maximum electric field associated to the produced EMPs was capable to exceed the value of 100 kV/m over a wide spectral band, covering the whole bandwidth of the scope used for the measurement (Lecroy 735 Zi, 3.5 GHz bandwidth) [22]. In Figure 6 the EMP spectra for typical shot conditions is reported. It was acquired by a D-Dot probe placed behind a 10 cm thick Teflon shield at 120 cm from the interaction point, where a cellulose triacetate (TAC, $C_{12}H_{16}O_8$) foam target (thickness = 425 µm, density 2 mg/cm$^3$) was irradiated with a first ns laser pulse followed, after 3 ns, by the 750 fs laser pulse having energy of 82.4 J. During this shot, the associated electric field estimated by the procedure reported in [23] reached ∼100 kV/m.

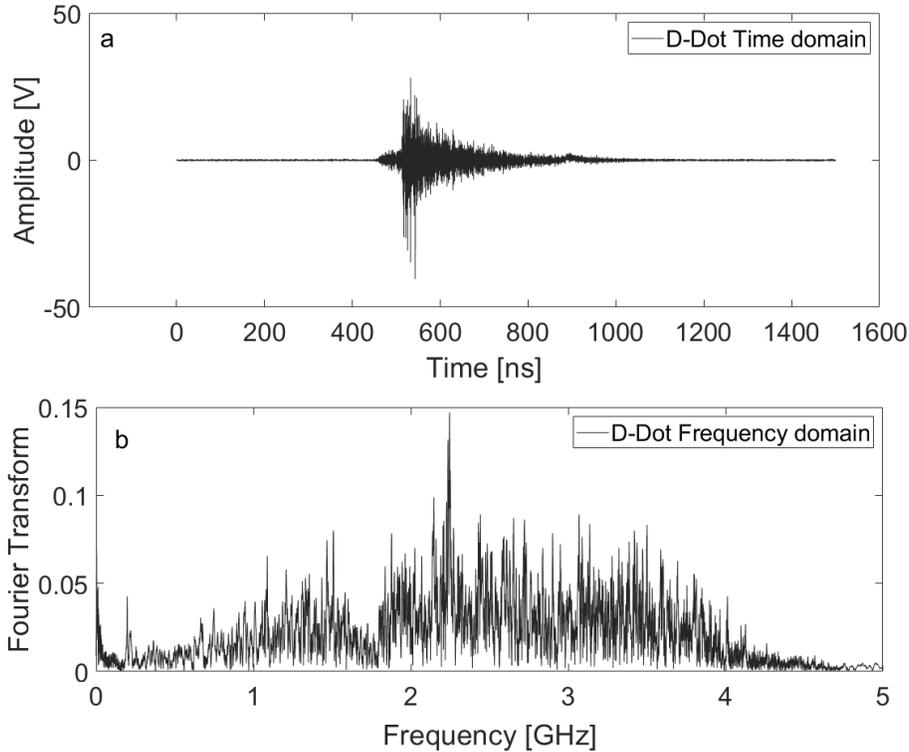

**Figure 6.** (a) The signal acquired by the D-Dot probe in the time domain; (b) The spectrum retrieved from the reported signal by suitable Fourier Transform.

During the same shot, the polycrystalline diamond detector provided the signal shown in Figure 7. The detector was biased at +300 V and placed on the horizontal plane at 90 cm from the interaction point at 37 degrees from the target normal axis. The signal provided by the diamond was collected by means of the Lecroy HDO 8108 scope (bandwidth = 1 GHz, ENOB = 8.6, sample rate = 10 GS/s). The time-of-flight line extension was made by means of an aluminium pipe with diameter of 40 mm, mounted on a KF-40 flange on the chamber wall. This enhanced the effective shielding to the produced EMP thanks to the waveguide action performed by the pipe itself [18]. Indeed it provided a cut-off frequency of 4.39 GHz and attenuated all the electromagnetic waves having higher frequencies [8,14,18]. During this shot the diamond detector was covered with a 20 µm thick aluminium filter. This allowed to cut heavier ion contribution at expenses of loss of information on the lower energy region of the proton spectrum (below ~1.2 MeV).

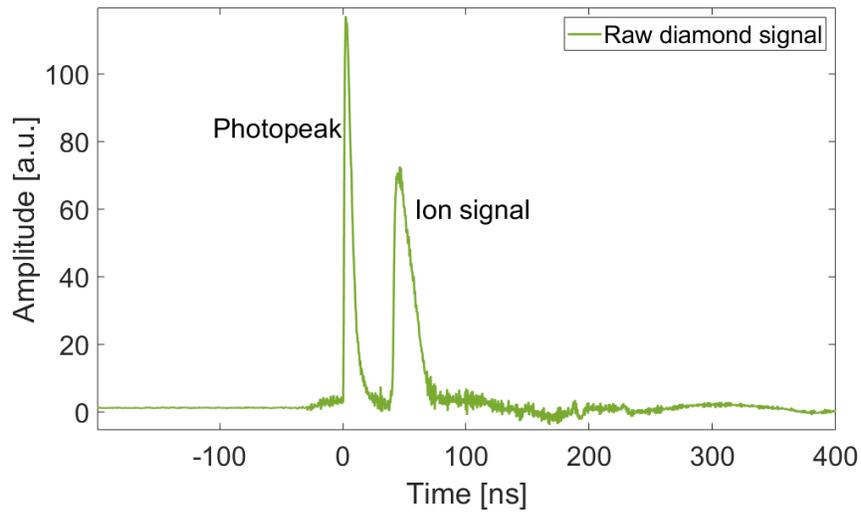

**Figure 7.** The raw-time domain signal collected by the polycrystalline diamond detector during the described shot.

The collected signal clearly presents a high signal to noise ratio. The temporal position of the photo peak is clearly recognizable as well as the instant when the protons with maximum energy are detected.

The calibrated proton spectrum, shown in Figure 8 together with the associated tolerances, was then retrieved by following the procedure described in [8]. To this purpose it was used the information on the detector temporal response and Charge Collection Efficiency that were obtained during the detector characterization, as shown in previous section.

The maximum measured proton energy was (2.2 ± 0.2) MeV and, as usual, the number of detected protons decreases for increasing energy.

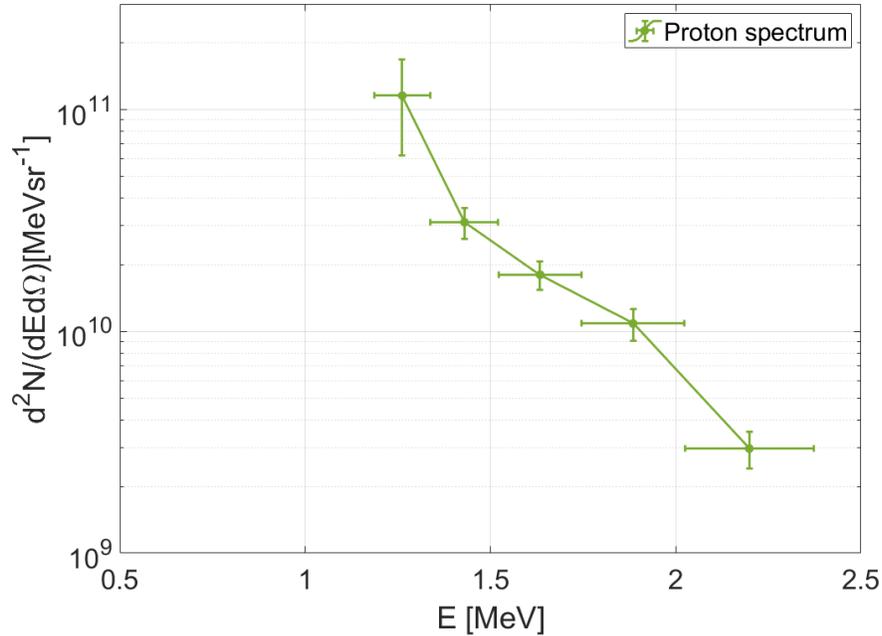

**Figure 8.** The proton spectrum obtained by the signal in Figure 7. The measured maximum proton energy is 2.2 ± 0.2 MeV.

As already mentioned, the information for energies lower than 1.2 MeV are lost due to the presence of 20 µm Al filter used for excluding the contribution of heavier ions. Nevertheless the feature of the spectrum for higher energies were retrieved with high accuracy. These particles loose part of their energy on the Al foil, and to reconstruct their accurate spectrum it is then necessary to make an estimation of the energy loss for each energy of the incoming particle. This was performed by means of SRIM [24] calculations, and the results for this attenuation are shown in Figure 9.
Here are also reported the attenuation value for a filter having 17 µm and 23 µm thicknesses, corresponding to the lower and upper limit of the 15% tolerance declared by the Goodfellow website, where the aluminum sheet was purchased.

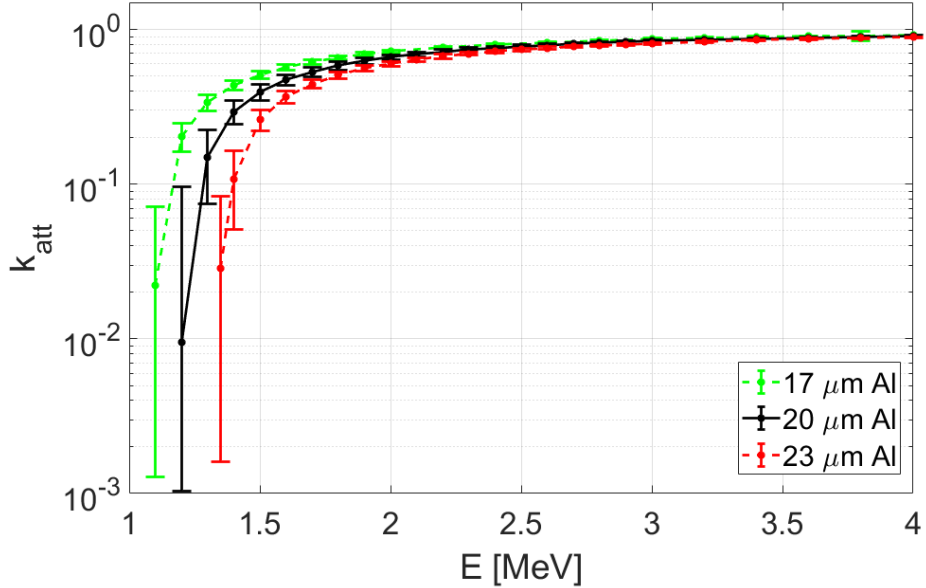

**Figure 9.** The Attenuation coefficient obtained by SRIM computation for the 20 µm aluminum filter together with the one obtained for 17 µm and 23 µm thicknesses corresponding to the tolerance of 15% declared by the aluminum sheet manufacturer.

## Conclusions

In the last decades the field of laser-matter interaction underwent a rapid expansion. The growing availability of powerful laser systems paved the way for the application of laser-driven ion sources to different scenarios, from inertial confinement fusion to material analysis for security and cultural heritage [25–28]. This growth needs to be accompanied by the development of suitable diagnostic systems able to effectively characterize the interaction. To do this, it is crucial to face some challenges, for instance to work at high repetition rate while maintaining a high sensitivity of the measurements, often hindered by the remarkable EMP fields typical of these scenarios.

In this work, we have described a novel detector design based on a 150 µm thick polycrystalline diamond sensor having a surface of 15 mm × 15 mm, characterized by a fast temporal response (FWHM = 3.47 ns) and a charge collection efficiency of ∼45%. This was employed in a TOF scheme during an experimental campaign held at the PHELIX laser facility at high pulse intensity and energy. This is recognized to be one of the main configurations where produced EMP fields are most intense [15] and in these conditions the detector assembly effectively proved its high EMP rejection. As a result it was thus capable of providing high sensitivity together with a fast temporal response. This allows for accurate real time characterization of ions accelerated by the laser-matter interaction, and thus for an effective related proton spectrum reconstruction.

The characteristics of this detector, in particular its high sensitivity, make it perfectly suitable to monitor processes characterized by low fluxes as, for instance, the α particles produced by laser-initiated p-$^{11}$B fusion reactions [13] or laser interactions with gas-jet targets. Moreover, thanks to the compactness and the easy handling of the detector structure it is also possible to position a set of them at different angles and distances from the interaction point. This would allow to obtain simultaneous measurement of the accelerated ions along different directions and to have a more complete reconstruction of the interaction process.

## Acknowledgement

The work has been carried out within the framework of the EUROfusion Consortium and has received funding from the Euratom research and training program 2014–2018 and 2019-2020 under grant agreement No. 633053. The views and opinions expressed herein do not necessarily reflect those of the European Commission.

The results presented here are based on the experiment P176, which was performed at the PHELIX facility at the GSI Helmholtzzentrum fuer Schwerionenforschung, Darmstadt (Germany) in the frame of FAIR Phase-0.

The research leading to these results has received funding from LASERLAB-EUROPE (grant agreement No. 654148, European Union's Horizon 2020 research and innovation program).

The research of N.E.A. was supported by the Ministry of Science and Higher Education of the Russian Federation (Agreement with Joint Institute for High Temperatures RAS No 075-15-2020-785, dated September 23, 2020).